\documentclass[twocolumn,showkeys,preprintnumbers,amsmath,amssymb]{revtex4}

\begin{document}
\bibstyle{apsrev}
\title{Quantum Computers, Discrete Space, and Entanglement}
\author{Mladen Pavi\v ci\'c}
\affiliation{University of Zagreb, 
GF, Kaciceva 26, POB 217, HR-10001 Zagreb, Croatia.}
\email{mpavicic@faust.irb.hr;  Web page: http://m3k.grad.hr/pavicic}


\date{\today}

\widetext
\begin{abstract}
We consider algebras underlying Hilbert spaces used by quantum information
algorithms. We show how one can arrive at equations on such algebras
which define $n$-dimensional Hilbert space subspaces which in turn
can simulate quantum systems on a quantum system. In doing so 
we use MMP diagrams and linear algorithms. MMP diagrams are tractable
since an $n$-block of an MMP diagram has $n$ elements while an $n$ block 
of a standard lattice diagram has $2^n$ elements. An immediate test for
such an approach is a generation of minimal and arbitrary 
Kochen-Specker vectors and we present a minimal  $n\ge 5$-dimensional 
``state-independent'' Kochen-Specker set of seven vectors. 

\end{abstract}



\maketitle

\section{Introduction}
\label{sec:intro}

In this paper we consider an algebra underlying Hilbert space 
used by quantum information algorithms and we explore whether 
one can use it for quantum computers in the same way one uses 
the Boolean algebra for classical computers. Since the 
answer is in the negative, we consider the possible modifications
of the afore mentioned quantum algebra which allow general 
applications of the algebra in formulating algorithms and 
simulating quantum systems. 

We will first present classical vs.~quantum algebras in the 
next section stressing that the quantum one has to be infinite. 
Since this is in contrast with finite quantum algebras 
available on quantum computers we proceed with presenting 
possible new finite quantum algebras and algorithms in 
Sec.~\ref{sec:algo}. As a result we obtain a general algorithm 
for obtaining Kochen-Specker vectors and therefore an 
automated proof of the Kochen-Specker theorem in Sec.~\ref{k-s}. 
   
\section{Algebras}
\label{sec:algs}

Classical computers standardly manipulate two-valued (0 and 1; 
bits, \it bi\/\rm nary dig\it it\/\rm s) elements of information 
using switches (physical devices) which are called gates. 
Their design is based on (a two-valued) Boolean algebra, also 
called \it switching\/ \rm algebra. 
We should stress here that a Boolean algebra based on $n$-valued 
elements is equivalent to the one based on two-valued elements.

A Boolean algebra is an algebraic structure consisting of a set
of elements together with two binary operations \it join\/\rm, 
$\cup$ and \it meet\/\rm, $\cap$ and a unary operation 
\it orthocomplement\/\rm, $^\perp$, such that the closure 
property holds, the law of distributivity, $a\cap(b\cup c)=
(a\cap b)\cup(a\cap c)$, associativity, and 
commutativity hold, and the identity and orthocomplement 
exist (so, it is a distributive lattice; a lattice is an ordered 
set in which all joins and meet exist). Each of the operations 
and their combinations can be implemented in the form of logic 
circuits by means of gates. 
Hence, one performs  a classical task by first digitizing 
it, then manipulating bits, and in the end translating bits back 
to the original language of the task (no classical computer can 
directly mimic a classical physical process). In doing so, one can 
access the values of all bits at any stage of their manipulation. 

Quantum computers  manipulate \it qubits\/ \rm (\it qu\/\rm 
antum \it bits\/\rm)---elements of quantum information (which 
are actually not digits but vectors (states) from Hilbert 
space) by means of \it quantum gates\/\rm.

Closed subspaces of a Hilbert space form an algebra called a Hilbert
lattice. A Hilbert lattice is an orthomodular lattice which, is by
definition a (relaxed) Boolean algebra in which the distributivity 
(see above) holds if $b\le a$ and $c\perp a$. 
In any Hilbert lattice the operation \it meet\/\rm, $a\cap b$, 
corresponds to set intersection, ${\cal H}_a\bigcap{\cal H}_b$, 
of subspaces ${\cal
H}_a,{\cal H}_b$ of a Hilbert space ${\cal H}$, the ordering relation
$a\le b$ corresponds to ${\cal H}_a\subseteq{\cal H}_b$, the operation
\it join\/\rm, $a\cup b$, corresponds to the smallest closed subspace of
$\cal H$ containing ${\cal H}_a\bigcup{\cal H}_b$, and
the \it orthocomplement\/ \rm $a^\perp$ corresponds
to ${\cal H}_a^\perp$, the set of vectors orthogonal to all vectors in
${\cal H}_a$. One can define all the lattice operations on a Hilbert 
space itself following the above definitions: 
${\cal H}_a\cap{\cal H}_b={\cal H}_a\bigcap{\cal H}_b$,
${\cal H}_a\cup{\cal H}_b=({\cal H}_a^\perp\bigcap{\cal
H}_b^\perp)^\perp$. Also, the orthogonality (mentioned above) 
${\cal H}_a\perp{\cal H}_b$ means ${\cal 
H}_a\le{\cal H}_b^\perp$.~\cite[p.~175]{isham}, 
\cite[pp.~21-29]{halmos}, \cite[pp.~66,67]{kalmb83},
\cite[pp.~8-16]{mittelstaedt-book}

Thus, using the properties of Hilbert space one arrives at a 
definition of the Hilbert lattice as an orthomodular lattice which 
satisfies: 

{\em Completeness:\/} The meet and join of any subset of
a Hilbert lattice always exist.

{\em Atomicity:\/} Every non-zero element in an {\em HL} is greater
than or equal to an atom. (An atom $a$ is a non-zero lattice element
with $0< b\le a$ only if $b=a$.)

{\em Superposition Principle:\/} (The atom $c$ is a superposition
of the atoms $a$ and $b$ if $c\ne a$, $c\ne b$, and $c\le a\cup b$.)
1. Given two different atoms $a$ and $b$, there is at least
one other atom $c$, $c\ne a$ and $c\ne b$, that is a superposition
of $a$ and $b$; 2. If the atom $c$ is a superposition of  distinct atoms
$a$ and $b$, then atom $a$ is a superposition of atoms $b$ and $c$.

{\em Minimal length:\/} The lattice contains at least
three elements $a,b,c$ satisfying: $0<a<b<c<1$.

One can also prove the other direction and therefore a Hilbert lattice 
is isomorphic to the set of closed subspaces of a Hilbert space.~\cite{maeda}
Here comes a result we want to stress: It can be proved that 
a Hilbert lattice \it must\/ \rm contain infinite number of 
atoms.~\cite{ivertsj} 
Moreover, if we wanted a Hilbert lattice to provide us with a complex 
field over which Hilbert space can be defined, we should assume 
that the Hilbert lattice contains a countable infinite sequence of 
orthogonal elements. 

Infinite dimensionality of a Hilbert space corresponds to 
the space continuity, to the integrals instead of sums, 
to radial functions and spherical harmonics, etc.; in a word, to 
all solutions of the Schr\"odinger equation we are used to. 
Therefore, the usual space distribution of, e.g., a wave function 
of electrons within, e.g., a molecule requires 
an infinite dimensional Hilbert space. Since we cannot have 
infinite dimensionality on a quantum computer we cannot directly 
simulate quantum mechanics on a quantum computer. But since 
both systems are quantum systems, a simulation---as opposed to 
the classical case---is nevertheless possible.   

\section{Algorithms}
\label{sec:algo}

In the literature simulation of quantum mechanics on a quantum 
computer has been approached in basically two ways. 
The first approach is to simulate one quantum system by another 
which resides in a quantum computer and might be simpler, e.g., 
proton spins by electrons in quantum dots.~\cite{lloyd,lloydc}  
Or even ``universal quantum computation over continuous variables 
for transformations that are polynomial in those 
variables.''~\cite{lloyd-braunst99}
This approach does not help us, though, since we have to find 
the algebra of the quantum computer system itself. 
The second approach is to simulate quantum mechanics by means 
of quantum gas model.~\cite{schr-simul,qgas} 
Basically this boils down to application of the 
corresponding Schr\"odinger equation on points in a grid.   
As a result the points sit in the grid so as to fit the 
continuous wave function. Hence, a discrete set of points 
approximates a continuous wave function but we still do not 
have a genuine discrete algebra and discrete Hilbert space. 
Our aim is to investigate whether such discretization is possible. 

We consider finite orthomodular lattices and filter them 
through the conditions stated in Sec.~\ref{sec:algs}  
and investigate properties which hold and which fail in 
the lattices. We want to find classes of 
such lattices which would approximate lattices with infinite 
number of atoms and in the end we want to compare them with 
lattices we derive from a finite dimensional Hilbert space. 

The most attractive feature of such a procedure is that one 
can define finite lattices by algorithms which are not simply 
read off from the standard Hilbert space properties but are 
derived form highly nontrivial theorems derived in the theory
of Hilbert lattices in the last 20 years. These algorithms 
also speed up calculations for several orders of magnitude. 
It can be shown that finite orthomodular lattices can be 
obtained from MMP diagrams which are organized as connected 
blocks of mutually orthogonal atoms. MMP diagrams are 
diagrams that are defined as follows: 

\begin{enumerate}
\item Every vertex (i.e., 
atom when a diagram corresponds to a lattice) belongs to 
at least one block; 
\item If there are at least two vertices 
then every block is at least 2-element; 
\item Every block which 
intersects with another block is at least 3-element;
\end{enumerate}
and then generated by the the isomorph-free generation 
procedure according to the following algorithm \cite{bdm-ndm-mp-1}: 
\begin{enumerate}
\item[]{\bf procedure} scan $(D${\rm : diagram;} $\beta${\rm : integer}$)$
\begin{enumerate}
\item[]{\bf if} $D$ {\rm has exactly $\beta$ blocks} {\bf then}
\begin{enumerate}
{\bf output $D$}
\end{enumerate}
\item[]{\bf else}
\begin{enumerate}
\item[]{\bf for} {\rm each equivalence class of extensions} $D+e$ {\bf do}
\begin{enumerate}
\item[] {\bf if} $e\in m(D+e)$ {\bf then} scan($D+e$,$\beta$)
\end{enumerate}
\end{enumerate}
\end{enumerate}
\item[]{\bf end procedure}
\end{enumerate}

Without the latter algorithm MMP diagrams would be nothing but 
Greechie diagrams~\cite{svozil-tkadlec} with one of the conditions 
dropped. The isomorph-free generation procedure is what make them 
very different. Greechie diagrams are a handy way to draw 
Hasse diagrams but Hasse diagrams get more and more intrinsically 
complicated when we enlarge the number of atoms. E.g., a 
four-atom Greechie block has 16 elements, a five-atom Greechie block 
has 32 elements, and an $n$-atom Greechie block has $2^n$ 
elements, so they soon become intractable. MMP diagrams 
are however just strings. A five vertex block has 5 elements, 
an $n$ vertex block has $n$ elements.  

Depending on parameters we use in their generation (parameters 
appear as options in our programs) 
MMP diagrams can be represented as lattices, but also as partially 
ordered sets, or as vectors from a Hilbert space which do not 
form a lattice; they can even be used for representing relations 
between vectors, planes, and subspaces of any $n$-dim space in 
classical physics. Which diagram will be appropriate for which 
purpose is determined by a selection procedure we use once they 
are generated. 

So, the 3 simple aforementioned conditions imposed on diagrams 
gives us all we need to get all finite lattices of arbitrary 
complexity: we just eliminate diagrams in which Hilbert lattice 
properties do not hold. We currently use programs 
which generate and use lattices with up to 100 atoms but for all 
results we have obtained so far, 15 to 28 atoms suffice. 

We were also able to reformulate Hilbert lattice properties and 
substitute 3 (conjectured all) classes of polynomial equations of 
the $n$-th order for the afore stated conditions.  
One such class was known before.~\cite{godow} And the other two, 
the orthoarguesian class of n-th order and \it quantum state 
equation\/ \rm class, we found only 
recently.~\cite{mpoa99,mporl00,mp02b} As the name of the 
latter class tells us, it is determined by the kind of states 
imposed on any Hilbert lattice of Hilbert space, i.e., 
by possible evaluation of Hilbert lattice elements. 

Let us see what makes a difference between a 
classical and a quantum state. 

A state on a lattice {\rm L} is a function 
$m:{\rm L}\longrightarrow [0,1]$
such that $m(1)=1$ and $a\perp b\ \Rightarrow\ m(a\cup b)=m(a)+m(b)$.
This yields $m(a)+m(a')=1$ and $a\le b\ \Rightarrow\ ((m(a)=1\ \Rightarrow
\ m(b)=1)$.

A nonempty set $S$ of states on {\rm L} is {\em classical\/} if

\noindent
$(\exists m \in S)(\forall a,b\in{\rm L})((m(a)=1\ \Rightarrow
\ m(b)=1)\ \Rightarrow\ a\le b)$

\noindent
and {\em quantum\/} if

\noindent
$(\forall a,b\in{\rm L})(\exists m \in S)((m(a)=1\ \Rightarrow
\ m(b)=1)\ \Rightarrow\ a\le b)$

Now we are able to prove the following 

\noindent {\em Theorem}. 
Any orthomodular lattice that admits classical states is a Boolean 
algebra.

\noindent {\em Theorem}. 
Any Boolean algebra admits classical states and 
any Hilbert lattice admits quantum states.  

\noindent {\em Theorem}. 
An orthomodular lattice that admits quantum states is 
still not necessarily a Hilbert algebra (lattice).

The proof of the latter theorem is simple: many of the 
orthoarguesian equations (characteristic of any 3 and more 
dimensional Hilbert space) fail in many MMP diagrams with loops 
of at least 5 blocks and interpreted as Hasse diagrams which allow 
quantum states. 

Taken together, we conjecture that an infinite dimensional 
Hilbert space can be 
represented by a polynomial quantum algebra (Hilbert lattice 
reformulated by means of orthoarguesian and state equations) of 
the $n$-th order with $n\rightarrow\infty$ and for finite $n$ 
such an algebra can be implemented on a would-be quantum computer.   
Qubits as the elements of the algebra obey superposition 
principle but do not allow a fixed evaluation (see above: there 
is no state for {\it every\/}  element of the algebra). 
This is due to particular way in which the orthogonality can 
be defined in MMP diagrams, i.e., in Hilbert lattices and 
Hilbert space, and this orthogonality turns out to be very 
promising in solving problems because it can be reduced to 
linear equations as opposed to the standard approach to the 
orthogonality which is nonlinear. The details are presented in 
the next section.    
 
\section{Orthogonality, entanglement, and Kochen-Specker theorem}
\label{k-s}

In 1993 ``[w]e propos[ed] a new experiment employing two independent
sources of spin correlated photon pairs. Two photons from different
unpolarized sources each pass through a polarizer to a detector.
Although their trajectories never mix or cross they exhibit
4th--order--interference--like correlations when the other two
photons interfere on a beam splitter even when the latter two do
not pass any polarizers at all''~\cite{p-s94,p-josab95}, 
independently of \cite{ben-tel,zukow-zeil93} and 
simultaneously with \cite{zukow-zeil93}. Later the obtained 
results have been verified ``experimentally ... [by] two pairs 
of polarization entangled photons and subject[ing] one photon from
each pair to a Bell-state measurement. This results in 
projecting the other two outgoing photons into an entangled 
state.''~\cite{pan-zeil} The very same scheme was also used for 
teleportation.~\cite{ben-tel,zukow-zeil93,p-s94,p-josab95,bow-zeil-tel,bow-zeil}

That ``entangle[ment] and correlat[ion] in polarization [of the 
other two photons] even when we do not measure polarization on 
the first two at all''~\cite{p-josab95} and also our discovery of 
a 100\%\ polarization correlation between unpolarized 
photons~\cite{p-pra94} we arrived at by investigating creation 
and annihilation operators when acting on orthogonal states 
in the second quantization formalism. 
We realized that this orthogonality is crucially 
different from the classical orthogonality. In entanglement 
we make a tensor product state and then extract just a part 
of the state. Since in the obtained Hilbert subspace the 
orthogonality of the one dimensional subspaces containing 
relevant vectors means that they are included in the span of 
the other one dimensional subspaces of the subspace, i.e., 
that the vectors are orthogonal to each other, span the 
considered subspace, and make it a Hilbert space. 

Exactly this property of the quantum orthogonality enables us 
to use linear instead of classical nonlinear equations. To see 
the meaning of the difference we will consider the old 
problem of finding finite  Kochen-Specker vectors which prove
the  Kochen-Specker theorem.    

Recently proposed experimental tests of Kochen-Specker theorem 
\cite{cabell-garc99,simon-zeil00} and 
disputes on feasibility of such experiments 
\cite{meyer99,kent99,mermin99,simon-zeil01,cabell-01} prompted 
a renewed interest in the theorem and this an additional reason 
for reconsidering the theorem.   

The original Kochen-Specker theorem \cite{koch-speck} produced 
a set of 117 3-dimensional Hilbert space vectors for which 
there is no way 
to assign 1's and 0's to their states and therefore no way to 
provide quantum space with a classical Boolean model. The proof 
was tedious and subsequent attempts to reduce the number of 
vectors gave the following minimal results: 33 \cite{peres} and 31 
\cite[p.~114]{peres-book} 3-dim vectors, 18 \cite{cabell-est-96a} 
and 14 \cite{cabell-est-96b} 4-dim vectors, 29, 31, and 34 5-dim, 
6-dim, and 7-dim vectors, respectively \cite{cabell-99},  
36 8-dim vectors \cite{kern-peres},  etc. Reducing the number 
of vectors turn out to be important because a direct connection 
between such vectors and an experimental setup can be 
established.~\cite{cabell-99} However, no general method for constructing 
sets of Kochen-Specker vectors has been proposed so far and here we
give one. 

The main idea of our approach is to first show that for particular set 
of orthogonal Hilbert space vectors one can impose no 0$\,$-1 state on the 
vectors. However, we do that using the Hilbert space orthogonality: 
$a\le b\cup c\cup$ \dots, not the standard one: $(a,b)=0$, $(a,c)=0$, 
\dots, which boils down to a non-linear system: 
$a_1 b_1+a_2 b_2+ a_3 b_3+\dots
=0$, $a_1 c_1+a_2 c_2+ a_3 c_3+\dots=0$, \dots But even this 
Hilbert orthogonality we do not ``calculate''---it is ``built in'' 
in the MMP diagrams by its generations algorithm. We only check whether 
one can or cannot impose classical 0$\,$-1 state on the diagrams. We 
then only have to find the one which does not allow such a state
and this is done by a simple program which follows the definition
of the classical state.     

In order to convince the reader we will find a minimal set of 
vectors from a 5 dim space as an example. The smallest MMP 
diagram we find not to allow a classical state 0$\,$-1 is 
the following one: $abc,cde,e\!f\!a,egb,dg\!f$ (where, e.g, $abc$ 
means an orthogonal triple: $a\perp b$, $a\perp c$, and $b\perp c$).  
Then we form equations corresponding to the inner products of 
5 dim orthogonal vectors being equal to 0 and solve the system. 
We deal with triples and not with quintuples since 
we only have to find a set which does not allow 0$\,$-1 valuation.
I.e., we follow the two Kochen-Specker (actually, Gleason's
\cite{zimba-penrose}) conditions: 
\begin{enumerate}
\item No two orthogonal rays are both assigned the value 1; 
\item In any group of $n$ mutually orthogonal rays, not all of 
the rays are assigned the value 0. 
\end{enumerate} 

Such triples are in principle just a part of a possible 
experiment. What is important is that for particular orthogonalities
between chosen vectors we cannot ascribe 0$\,$-1 values to them. 

Since only the directions of the vectors (``rays'' \cite{peres}) are 
relevant they must be real. Since we did not care to find ``nice 
looking'' vectors some vectors are ``big'' due to a recursion 
procedure. This however do not affect the main aim of finding 
the vectors and it is to show that our approach works. 
a=\{608683911, 17315878, -22061625, -111556858, 20961326\}, 
b=\{3, 68, -123, 52, 4\}, 
c=\{1, 3, 5, 7, 11\}, 
d=\{11, -11, 11, -11, 4\}, 
e=\{1788, -8663, -1348, 8223, -2420\}, 
f=\{5791304343, -304905182408,\break-1387655556967, 1686769435032, 7600253389432\}, 
g=\{1, 1, 1, 1, 0\}. 
The reader can introduce the vectors into, e.g.,  Wolfram's \it Mathematica\/ 
\rm and convince her- or himself that they really are orthogonal and, 
by simple combinatorics, that one cannot ascribe 0 or 1 to all of them
(in each triple one element must be 1 and the other two zero and 
this is impossible).    

Cabello \cite{cabell-99} related his Kochen-Specker set of 18 vectors 
in 9 blocks with his experimental proposal in a four-dimensional 
Hilbert space~\cite{cabell-garc99} 
and he deals with 4-tuples. We deal with triples and leave a problem 
of finding  a related experiment open. This is because we are first of
all interested in finding a general algorithm for find all orthogonalities 
that do not allow 0$\,$-1 states. So, e.g., Cabello's 18 vectors in 9 blocks
can form 1430 MMP diagrams that do not allow 0$\,$-1 states. 
but do allow quantum states. Still, none 
of these examples (therefore not even the one elaborated in 
Ref.~\cite{cabell-99}) by itself correspond to a Hilbert space 
because their MMP diagrams do not correspond to lattices: 
the smallest triple lattice which do not allow states are two lattices 
with 19 atoms and 13 blocks and a quadruple lattice can only have 
more atoms and/or blocks. Other smaller cases with 18 vectors 
which do not allow 0$\,$-1 states are: 4 diagrams with 8 (quadruple 
blocks) blocks. The lowest number of 
quadruple blocks and vectors are: 1 4-block case with 10 vectors. 
 
Let us be more specific: one of the obtained 18-9 MMP diagrams is: 
$abcd$, $de\!f\!g$, $ghij$, $jklm$, $mnop$, $pqra$, $bikr$, $celn$,
$\!f\!hoq$.
And with $a$=100\=1, $b$=0110, $c$=11\=11, $d$=1\=111, $e$=111\=1,
$f$=0101, $g$=10\=10, $h$=010\=1, $i$=1\=11\=1, $j$=1111, 
$k$=11\=1\=1, $l$=1\=100, $m$=001\=1, $n$=0011, $o$=1000, $p$=0100,
$q$=0010, $r$=1001, 
this is nothing but Cabello's 18-9 case. Graphically it means 
a hexagram with 3 ellipses contained in it. The smallest 4-block 10-5 
case is: $abcd$, $de\!f\!g$, $ghia$, $bfij$, $cehj$. Graphically it 
means a triangle with 2 ellipses contained in it (with one common 
vertex not contained in the triangle). However, it might turn out 
(we still have not checkedt) that so small a diagram cannot be ascribed 
real vectors in a 4-dim space and that we should go to higher 
dimensions to find real vector sets. 
 
 \section{Conclusion}
\label{concl}

We have shown that one can build an algebra underlying Hilbert space 
which could be a universal algebra for quantum computers in the same 
way the Boolean algebra is for classical computers. 
In our approach the algebra is based on polynomial series of 
relations between one dimensional subspaces of a Hilbert space 
and linearly defined orthogonality relations between either 
subspaces or vectors of a Hilbert space. 

Linear orthogonality defined through MMP diagrams possibly opens 
a way to substitute a linear for nonlinear coupling between qubits 
presently required for universal quantum computation. On the 
other hand such linear orthogonality defined through MMP diagrams
already on our classical computers enabled speeding up calculations 
for more than 5 orders of magnitude on the CPU time scale and 
enabled us to find polynomial expressions of the $n$-th order 
representing any Hilbert space and unknown so far. It also 
enabled us to find a general approach to finding Kochen-Specker 
vectors, some of which we presented above.

\parindent=20pt
\acknowledgments
The author acknowledges support of the Ministry of Science of Croatia.

\end{document}